\documentclass[aps,pra,floatfix,twocolumn,showpacs]{revtex4}
\usepackage{dcolumn}

\begin{document}
\draft
\title{Resonance reactions and enhancement of weak interactions in collisions of cold molecules}
\author{V.V. Flambaum and J.S.M. Ginges}
\address{School of Physics, University of New South Wales, 
Sydney 2052, Australia}
\date{\today}

\begin{abstract}

With the creation of ultracold atoms and molecules, a new type of 
chemistry - ``resonance'' chemistry - emerges: chemical reactions can occur when 
the energy of colliding atoms/molecules matches a bound state 
of the combined molecule (Feshbach resonance). 
This chemistry is rather similar to reactions that take place in nuclei at low energies. 
In this paper we suggest some problems for  
future experimental and theoretical work related to the resonance chemistry of 
ultracold molecules.
Molecular Bose-Einstein condensates (BECs) are particularly interesting because in this system 
collisions and chemical reactions are extremely sensitive to weak fields; 
also, a preferred reaction channel may be enhanced due to a finite number of final states. 
The sensitivity to weak fields arises due to the high density of narrow compound resonances
and the macroscopic number of molecules with kinetic energy $E=0$ 
(in the ground state of a mean-field potential).
The high sensitivity to the magnetic field may be used to measure the 
distribution of energy intervals, widths, and magnetic moments of compound resonances and study the
onset of quantum chaos.
A difference in the production rate of right-handed and left-handed chiral molecules 
may be produced by external electric ${\bf E}$ and magnetic ${\bf B}$ fields and the finite width $\Gamma$
of the resonance (correlation $\Gamma {\bf E}\cdot {\bf B}$).
The same effect may be produced by the parity-violating energy difference in chiral molecules.

\end{abstract}
\vspace{1cm}
\pacs{03.75.Nt, 82.30.Cf, 32.80.Ys, 33.55.Ad}

\maketitle

\section{Introduction}

Techniques to produce and manipulate ultracold atoms and molecules are rapidly developing. 
Bose-Einstein condensation (BEC) has been realized in 
numerous dilute atomic gases \cite{firstBEC}
and, recently, in molecules \cite{molBEC}.
In these systems a new type of chemistry - ``resonance chemistry'' - emerges. 
Chemical reactions can take place when the energy of the free atoms/molecules 
matches the energy of a bound state of the combined system 
due to resonant coupling between the free and bound states.
The relative energies of the states can be tuned by variation of a magnetic field.
Ultracold molecules have been formed from degenerate Fermi gases \cite{regal}
and atomic BECs \cite{durr}
at Feshbach resonances \cite{tiesinga,inouye}, 
where two free particles resonantly couple to a bound state of the combined system.  
Just recently, the resonant coupling of three free atoms to a bound state of a triatomic molecule 
(Efimov resonance) has been observed \cite{kraemer}.
The aim of this short paper is to attract attention to some interesting 
problems related to resonance chemistry in ultracold molecules. 
Some results have already been presented in our preprint \cite{FG}.

\section{Dense spectrum of compound states, enhancement of weak perturbations, 
and quantum chaos}

The collisions of atoms in a BEC can be controlled by the variation of 
a relatively weak magnetic field; the magnitude and sign of the 
scattering length can be changed by varying the field near a Feshbach 
resonance \cite{stwalley,tiesinga}. 
In this way the expansion and collapse of an atomic BEC 
has been observed \cite{bosenova}. 
In molecules, a change in the scattering 
length can be achieved with a magnetic field much weaker than 
that used to obtain the same effect in atoms. 
This increase in the sensitivity to weak fields is due to the much richer 
spectra of molecules: there is an exponential 
increase of the density 
of resonances with the number of ``active'' particles.

Consider, for example, the formation of an intermediate non-stationary
compound state of four atoms arising from the collision of two cold diatomic
molecules. This state may be considered as an excited
state of the four-atom molecule (if this molecule has a finite binding
energy).
Compound states in molecules combine many electronic, 
vibrational, rotational, and hyperfine (spins of nuclei) degrees of freedom. 
Naturally, the interval between these compound levels is very small. 
Even without nuclear spins, the density of states may exceed  $10^5$ states per 
1~K energy interval ($2 \times 10^5$~K$^{-1}$ for collision of two PbO molecules  \cite{PbO}).
Simultaneously, the decay width
of these complex compound states decreases since the emitted
particle (atom or electron) must collect the energy from many
 degrees of freedom to have enough energy to escape from a potential well
produced by the interaction with other particles forming the compound state.
Similarly, in radiative decay one electron must collect
energy from all degrees of freedom to radiate an energetic photon.
This phenomenon of exponential decrease of the intervals between
energy levels and decrease of the decay widths is well-known in nuclear physics. 

Determination of the energy intervals, widths, and magnetic moments
of resonances in cold molecular collisions is an interesting problem by itself and may help
to find out if there is ``quantum chaos'' in this problem 
(a discussion of chaos in molecular collisions may be found, for example, in 
Ref. \cite{PbO} and references therein).
Conventional signatures of chaos include Wigner-Dyson statistics
of energy intervals between compound states, Porter-Thomas
statistics of capture and decay widths, statistics
of magnetic moments, etc. (see, e.g., the review \cite{Lane}).

Indeed, the spectra and wave functions of compound states in nuclei are usually chaotic.
This is because the residual interaction between the particles
exceeds the interval between the energy levels. Therefore, the
wave function of a compound state becomes a superposition
of a huge number of basic components (Slater determinants built
from products of single-particle orbitals). The expansion coefficients
in these linear combinations of basic components behave like
``random variables''. Quantum chaos also appears in excited
states of some rare-earth and actinide atoms and in many multiply-charged ions
\cite{Gribakin}.
In cold molecular collisions, compound states should also be chaotic superpositions
of a large number of basic components (a basic component here
is a state which may be classified using certain hyperfine, 
rotational, vibrational, and electronic quantum numbers).

In nuclei the statistics pertaining to quantum chaos are obtained
by changing the energy of the initial particles.
 In cold molecular collisions the statistics
may be studied by applying a magnetic field which changes the positions
of the compound resonances at a given energy.
 A very interesting manifestation
of quantum chaos is the enhancement of weak interaction effects
(see, e.g., reviews \cite{sushkov} and references within).     

\section{Difference in production rates of chiral molecules}

\subsection{Parity violating weak interaction}

The high sensitivity of ultracold molecules 
to weak fields could be used to search for an 
energy difference between chiral molecules produced by the 
parity violating (PV) weak interaction. As we will see, this 
difference leads to a difference in the production rate of right-handed 
and left-handed chiral molecules.
The interest in these problems with chiral molecules is mainly motivated by attempts to
 find a mechanism which produced homo-chiral biological molecules 
and to obtain a better understanding of biological evolution.    
It is well-known that biological molecules have a definite chiral structure 
(for example, there are only naturally occurring left-handed 
amino acids and right-handed sugars) \cite{pvreviews}. 
There have been numerous attempts to explain this effect 
by the influence of the PV weak interaction, 
which breaks the energy equivalence of right- and left-handed molecules 
\cite{rein,letokhov}. 
That parity violation can discriminate between molecules of different 
chirality is easily seen:
the PV electron-nucleus interaction 
in atoms \cite{khriplovich,bouchiats} creates a spin helix of the 
electrons which interacts differently with right- and left-handed molecules.
However, the parity violating energy difference $\Delta E_{PV}$ 
is very small \cite{RHS}, 
\begin{equation}
\label{pved}
\Delta E_{PV}\sim 10^{-20}Z^{5}\eta~{\rm a.u.} \ ,
\end{equation} 
where $Z$ is the nuclear charge of the heaviest atom, 
and $\eta$ is an asymmetry factor which can be found from 
molecular structure calculations. This strong dependence on $Z$ originates
from the weak ($\propto Z^3$) and spin-orbit ($\propto Z^2$) interactions.
It may appear that in molecules with heavy atoms $\Delta E_{PV}$ could 
become relatively large due to the $Z^{5}$-dependence. 
However, the asymmetry factor $\eta$ remains very small.
The effect may be orders of magnitude larger for molecules with two 
heavy atoms \cite{RHS}. 
For calculations of $\Delta E_{PV}$ for various molecules, 
see, e.g., \cite{molcalcs} and references therein. 
So far a PV energy difference in molecules has eluded 
 detection (see, e.g., \cite{det}).

Let us consider how a PV energy difference could 
manifest itself in the collision of two ultracold molecules.
Remember that in order to form a chiral molecule there must be at 
least four atoms involved \cite{four}; therefore, the collision of two diatomic 
molecules is sufficient.
The cross-section for formation of a chiral compound 
molecular state due to $s$-wave scattering can be expressed, 
using the Breit-Wigner formula, as 
\begin{equation}
\sigma = \frac{\pi}{k^{2}}\frac{\Gamma _{c}\Gamma}
{(E-E_{0})^{2}+\Gamma ^{2}/4} \ ,
\end{equation}
where $k$ is the wave vector, $\Gamma _{c}$ 
is the capture width, and $\Gamma$ is the total width of the 
resonance.
The PV weak interaction in the chiral molecules shifts the 
resonance energies. For example, let's consider that for the left-handed 
structure $E_{0}\rightarrow E_{L}=E_{0}-\Delta E_{PV}/2$ while for the 
right-handed structure 
$E_{0}\rightarrow E_{R}=E_{0}+\Delta E_{PV}/2$.
Therefore, cross-sections for the formation of left 
and right molecules, $\sigma _{L}$ and $\sigma _{R}$, from 
achiral components may be different. 
We can define an asymmetry parameter
\begin{equation}
P=\frac{\sigma_{R}-\sigma _{L}}{\sigma _{R}+\sigma _{L}} \ .
\end{equation}
The maximum value for $P$ is reached when $E=E_{0}\pm \Gamma/2$. 
At this energy the asymmetry parameter
\begin{equation}
\label{Pmax}
|P_{\rm max}|= \frac{\Delta E_{PV}}{\Gamma} \ .
\end{equation}
In principle, the resonances can be shifted  to
the point of maximum $P$, 
$E_{0}\pm \Gamma/2 \approx 0$, by application of an external 
electric or magnetic field.

A difference in the number of right-handed and left-handed 
chiral molecules, proportional to the asymmetry parameter $P$, produces optical activity. 
This may be a method of detection of the PV effect.
Note that PV experiments with atoms have already
demonstrated a very high sensitivity to small angles of rotation of the light polarization plane 
\cite{khriplovich,bouchiats}.
One may also try to detect the circular polarization
of light emitted in the decay of compound states
of chiral molecules. Indeed, light emitted by a chiral molecule has a 
certain degree of circular polarization. Therefore, a small difference 
in the number of right-handed and left-handed molecules produces some very 
small circular polarization.

Let us briefly consider what we may expect for the size of the optical rotation.
The angle of rotation $\varphi$ of the polarization plane of light passing through the sample 
may be expressed in terms of the asymmetry parameter $P$ as 
\begin{equation}
\varphi = \frac{N_R}{N_R+N_L}\varphi _R + \frac{N_L}{N_R+N_L}\varphi _L = P \varphi _R \ ,
\end{equation}
where $N_R$, $N_L$ are the numbers of right- and left-handed molecules and $\varphi_R$, $\varphi_L=-\varphi_R$ 
are the angles of rotation that the respective molecules produce. 
We use Eq. (\ref{Pmax}) to estimate $P$. 
The largest values for $\Delta E_{PV}$ that have been considered in molecular 
calculations are $\sim 10^4 {\rm Hz}$ 
(e.g., for H$_{2}$Po$_{2}$ \cite{molcalcs}).
The width of the level $\Gamma$ may be quite small, since the capture
width $\Gamma_c=0$ for energy $E=0$ and   
the radiative width may be much smaller than $10^6 {\rm Hz}$
(a typical width for optical photon emission) because of 
``chaotic''  suppression discussed at the beginning of this paper. 
One should check if there are additional decay channels: the four-atom
compound state may have enough energy to decay by emission of one atom or electron.
The actual value of the width, which also includes a width due to collisions, 
should be determined experimentally. 
In $^{133}{\rm Cs}$, very narrow widths, as small as ${\rm 3.5\ kHz}$, have been observed in 
$g$-wave Feshbach resonances \cite{herbig}.
Therefore, an asymmetry parameter $P\sim 1$ is not out of the question.  
The value for the rotation angle $\varphi _R$ depends on a number of parameters: the molecular density, 
the refractive index, the pseudoscalar polarizability, the wavelength of the light, and the pathlength 
\cite{condon,pola}.
A well-known example of a chiral molecule is sucrose; in water solution, 
it rotates light at the sodium D-line by an angle $\sim 700\ \rho \ \deg /{\rm m}$, 
where $\rho$ is the density (or concentration) in ${\rm g}/{\rm cm}^3$. 
The numerical coefficient is an intensive property of the molecule, depending on the temperature of 
the sample and the wavelength of light. The value for sucrose is not untypical. 
We can expect similar values for chiral molecules in gases, i.e. a rotation angle $\sim 1 \ {\rm deg}/{\rm m}$. 
In a very low-density gas the angle of rotation can be significantly increased by tuning the light frequency to resonance.

In collisions in molecular BECs there may be some enhancement of the effect due to the 
macroscopic number of molecules in the ground state with vanishingly small energy spread. 
Moreover, the effect may be Bose-enhanced (non-linear 
enhancement of the reaction due to the number of identical particles in the final state). 
Through a small difference in reaction rates such enhancement can lead to an almost complete 
selectivity of one reaction channel over another \cite{moore}. 
This could be a mechanism for selectivity of chiral molecules of one handedness produced, e.g., 
by a parity violating energy difference or the fields ${\bf E}\cdot {\bf B}$ (see the following section).

\subsection{Pseudoscalar correlation produced by electric and magnetic fields}

Constant homogeneous fields cannot produce an energy difference between
molecules of different chirality. Indeed, 
a chiral molecule is characterized by a pseudoscalar
$[{\bf n_A}\times {\bf n_B}] \cdot{\bf n_C}$, where ${\bf n_A}$, 
${\bf n_B}$, and ${\bf n_C}$ are vectors showing locations
of atoms $A$, $B$, and $C$ relative to atom $D$.
The weak interaction is proportional to a pseudoscalar ${\bf s} \cdot{\bf p}$, 
where ${\bf s}$ and ${\bf p}$ are the electron spin and momentum.
This interaction produces chirality-dependent energy shifts
since its pseudoscalar ${\bf s} \cdot{\bf p}$ may be correlated
with the molecular pseudoscalar $[{\bf n_A}\times {\bf n_B}] \cdot{\bf n_C}$
(with the help of the spin-orbit interaction which links the spin
${\bf s}$ with the coordinate variables).
To imitate the PV energy difference
we need to make a $T$-invariant pseudoscalar (effective interaction)
from electric and magnetic
fields. All combinations like  ${\bf E}\times {\bf B}$,  ${\bf E}\cdot {\bf B}$,
${\bf E}\cdot {\bf E}$, and ${\bf B}\cdot {\bf B}$ do not satisfy
this requirement. The molecular vectors ${\bf n_A}$, 
${\bf n_B}$, and  ${\bf n_C}$ cannot be included into these
effective interactions since they disappear after averaging over
molecular orientations.
In principle, one can make the $T$-even pseudoscalar by considering
inhomogeneous and time-dependent fields
($ {\bf B} \cdot \frac{\partial{\bf E}}{\partial t} $,
${\bf E} \cdot \frac{\partial{\bf B}}{\partial t} $, etc.).
However, the corresponding effects should be very small
(proportional to the change of the field on the molecular scale). Therefore,
stray fields can hardly imitate effects of the PV weak interaction.

The only exception here may be the correlation ${\bf E}\cdot {\bf B}$.
It violates both $P$ and $T$ invariance, therefore it cannot produce an energy
shift of a stationary state. However, a compound resonance is not a
stationary state. Therefore, this correlation can induce a difference
in the production of right-handed and left-handed chiral molecules due to
the finite width of the compound state or due to any final state interaction
in general. In principle, this may be a natural source of asymmetry between biological 
chiral molecules.
A reliable way to find the magnitude of the effects is to perform a dedicated 
experiment with controlled fields.

\subsubsection{Origin of the  $T$-odd correlation in circular polarization of photons}

It is important to show how the $T$-odd correlation ${\bf E}\cdot {\bf B}$ 
can produce $T$-even effects. We will consider a simpler effect 
produced by these fields: 
circular polarization of photons in atomic transitions.
This possibility was first pointed out for hydrogen in Ref. \cite{Azimov}.

The magnetic field alone can produce circular polarization for photons
emitted in a definite direction. This effect has the same origin as 
Faraday rotation and can be described by the correlation 
${\bf B}\cdot {\bf k} \lambda$, where ${\bf k}$ is the unit momentum vector and 
$\lambda$ is the helicity (photon states with definite circular polarization 
correspond to $\lambda =\pm 1$).
Such circular polarization disappears after averaging over photon directions.
The case of two fields with non-zero ${\bf E}\cdot {\bf B}$ is different.
Here the circular polarization does not vanish after averaging.
Indeed, the helicity (circular polarization) of a particle $\lambda={\bf s}\cdot {\bf k}$
is a $T$-even pseudoscalar (${\bf s}$ is the unit spin vector).
As known in atomic transitions, 
a small photon circular polarization is normally produced by the weak interaction
which is also a $T$-even pseudoscalar. The correlation ${\bf E}\cdot {\bf B}$
is a $T$-odd pseudoscalar, therefore it may create circular polarization
proportional to the widths of the involved quasistationary states or any 
final state interactions. Thus, the situation with the production
of helicity/circular polarization is similar to the
situation with the production of chirality.

The result for the circular polarization $P$ can be schematically
presented using the following notations. Let us assume that
an atomic electron is excited from a ground state to a non-stationary
state $p_{1/2}$ with energy $E_{1/2}$, width $\Gamma_{1/2}$, and
excitation amplitude $T_{1/2}$. Then this state decays to some final state $f$ 
and emits a photon with real radiation amplitude M1. The amplitude of this 
process can be presented as 
\begin{equation}
\label{A}
A=\frac{T_{1/2} M1}{E-E_{1/2}+i\Gamma_{1/2}/2} \ .
\end{equation}
Here $E$ is the excitation energy. We assume that $E \approx E_{1/2}$. 
Assume that close to the $p_{1/2}$ state
there are states $p_{3/2}$ and $s_{1/2}$. The state $s_{1/2}$
may decay to the same final state with imaginary radiation amplitude $i E1$
(origin of the imaginary unit $i$ in this formula may be found, e.g., 
in the book \cite{khriplovich}).

The combined effect of the magnetic and electric field gives us another amplitude:
\begin{equation}
\label{B}
B=\frac{T_{1/2}\langle p_{1/2}|\mbox{\boldmath$\mu$} \cdot {\bf B}|p_{3/2}\rangle 
\langle p_{3/2}|e{\bf r}\cdot {\bf E}|s_{1/2}\rangle iE1}{(E-E_{\frac{1}{2}}+\frac{i\Gamma_{\frac{1}{2}}}{2})
(E-E_{\frac{3}{2}}+\frac{i\Gamma_{\frac{3}{2}}}{2})(E-E_{s}+\frac{i\Gamma_{s}}{2})} \ .
\end{equation}
We may assume here that both fields ${\bf B}$ and ${\bf E}$ are directed 
along the $z$-axis. 
Then summation over the angular momenta of the intermediate states gives 
a trivial positive coefficient.

For comparison we present here the amplitude $B^W$ produced by the
weak interaction:
\begin{equation}
\label{B^W}
B^W=\\
\frac{T_{1/2} \langle p_{1/2}|W|s_{1/2} \rangle
 iE1}{(E-E_{1/2}+i\Gamma_{1/2}/2)
(E-E_{s}+i\Gamma_{s}/2)} \ .
\end{equation}
In this  case $B$ is induced
by the imaginary weak matrix element $\langle p_{1/2}|W|s_{1/2} \rangle=iW$.
The calculation of the photon circular polarization $P$ arising from the weak interaction 
is presented in \cite{khriplovich}.

The circular polarization $P$ appears due to the interference of the amplitudes $A$ and $B$.
The relative sign of the $E1$ and $M1$ amplitudes depends on the photon circular
polarization. One may schematically present this dependence
in the total amplitude as $M_{\pm}=A \pm B$, where ``$+$''
 and ``$-$'' correspond to the right-handed and left-handed circular
polarizations. As a result, the average circular polarization is 
\begin{equation}
\label{P}
P \sim \frac{AB^*+A^*B}{|A|^2 + |B|^2} \ .
\end{equation}

In the case of the weak interaction the factor $i$ in the amplitude
$iE1$ is compensated by the similar factor in the weak matrix
element $iW$. In our case the situation is different:
if all widths are zero, $AB^*+A^*B=0$. Thus, the result is proportional
to these widths. This difference is explained by the fact that the
weak interaction and the circular polarization are time reversal
$T$-even  while the correlation ${\bf E}\cdot {\bf B}$
is $T$-odd. Therefore, we need widths which manifest a time asymmetry
of the problem. 

This model calculation shows us how widths appear in the  
${\bf E}\cdot {\bf B}$ effect for chirality production.
It also gives us a very rough estimate for the magnitude of the effect:
\begin{eqnarray}
\label{M}
P &\sim& D \frac{\Gamma \mu ea_B {\bf B}\cdot {\bf E}}
{(E_{1}-E_{3})^2(E_{1}-E_{2})} \eta \ \\
&\sim& D' [{\bf B}/{\rm T}] \cdot [{\bf E}/(10^{4}{\rm V/cm})] \ , 
\end{eqnarray}
where $D$ is a numerical coefficient that takes into account (chaotic) suppression of the 
magnetic and electric dipole matrix elements compared to the Bohr magneton $\mu$ and Bohr radius $a_B$ 
(however, this suppression is more than compensated by a larger enhancement from 
the very small molecular energy denominators in Eq. (\ref{M}), see Ref. \cite{sushkov}) and  
$\eta$ is an asymmetry factor which, as experience with the weak interaction shows, may be quite small.
The value for $D'$ (overall numerical coefficient depending on $D$, $\eta$, and energy intervals and widths) 
strongly depends on the molecule under consideration and 
can only be reliably determined from 
experiment or sophisticated molecular structure calculations which are beyond the scope of this work.
There is a very dense spectra of Feshbach resonances in collisions of polar molecules; 
taking modest values for widths and energy intervals from Ref. \cite{PbO}, allowing for suppression of matrix elements, 
and choosing a value $\eta \sim 10^{-3}$ for the asymmetry factor (see, e.g., Ref. \cite{molcalcs}), 
a rough estimate gives $D'\sim 10^{-7}$. 
The true coefficient for a molecule with dense spectra could be several orders of magnitude larger.

\section{$P$-even and $P$-odd correlations in collisions of cold molecules}

In principle, PV and ${\bf E}\cdot {\bf B}$
circular polarizations in the decay of compound
states formed in the collision of cold molecules can be measured. 
Here we can even have a certain enhancement
in comparison with the circular polarization in atoms because
of close levels of opposite parity in the compound spectrum  - see the small 
energy denominators in Eqs. (\ref{A},\ref{B},\ref{B^W},\ref{P}).
Moreover, in a BEC, Bose-enhancement may lead to almost complete selectivity of the circular polarization 
of emitted photons due to amplification of the decay of the compound state through the preferred channel 
(the projection of the angular momentum in the final state of the molecule is different 
for different circular polarizations of the emitted photon).

There is another manifestation of parity violation in resonance collisions 
of cold atoms or molecules. It is related to the admixture of an 
$s$-wave to a $p$-wave compound resonance. 
For energy $E=0$, only $s$-wave molecules have a significant 
interaction cross section. Consider now a $p$-wave compound resonance. 
It seems to be invisible for $E=0$. 
(Note that the $p$-wave amplitude is actually not exactly zero
since the trap potential and mean field make the kinetic energy
$E$ non-zero even in the ground state.) 
The weak interaction $W$ mixes states of opposite parity 
and produces the combined state $|\psi\rangle=|p\rangle +\beta|s\rangle$, 
thus opening the $s$-wave reaction amplitude proportional to $\beta$. 
The mixing coefficient  $\beta=\langle p | W |s\rangle/(E_s-E_p)$ 
is enhanced since the energy interval between the opposite parity 
compound states $(E_s-E_p)$
is very small due to the high level density in a combined molecule.
A similar mechanism is responsible for the enhancement of 
weak interaction effects in neutron-nucleus reactions $\sim 10^{6}$ 
times \cite{sf} (for a review of the experiments, see \cite{nNexp}).
Interference of the very small $p$-wave amplitude 
and the weak-induced $s$-wave amplitude leads to PV effects
proportional to \cite{sf,ngamma}
\begin{equation}
P=\sum_s\sqrt{\frac{\Gamma_s(E)}{\Gamma_p(E)}}
 \frac{i \langle p | W |s\rangle}{(E-E_s)} \ ,
\end{equation}
where $\Gamma_p(E)$ and $\Gamma_s(E)$ are capture widths for
the $p$-wave and nearby $s$-wave resonances taken at the actual collision
energy $E$ which is assumed to be close to $E_{p_{1/2}}$. 
The kinematic enhancement factor
 $\sqrt{\Gamma_s(E)/\Gamma_p(E)}=T_s/T_p \sim 1/kR$ tends to infinity
at small energies; here
$k=p/\hbar \propto  \sqrt{E}$ is the molecular wave vector,
 $R$ is the size of the molecule, $T_s$ and $T_p$ are $s$-wave and
$p$-wave capture amplitudes. 
PV effects produced in this way involve correlations
like ${\bf S} \cdot{\bf p}$ or 
 $\Gamma [{\bf n_A}\times {\bf n_B}]  \cdot{\bf p}$, 
where $S$ and $p$ are molecular spin and momentum. There is also a number of parity
conserving correlations in the photon emission process,
similar to that studied in nuclear reactions with compound resonances
\cite{ngamma}. An even richer picture appears in molecules
since here we may consider parity violating and parity conserving correlations 
produced by external fields ${\bf E}$ and ${\bf B}$.

\acknowledgments

We are grateful to D. Budker and J. Higbie for useful discussions.
This work was supported by the Australian Research Council.


\begin{thebibliography}{20}

\bibitem{firstBEC}

For the first observations of BEC in gases, see: 
M.H. Anderson {\it et al}., Science {\bf 269}, 198 (1995); 
C.C. Bradley {\it et al}., Phys. Rev. Lett. {\bf 75}, 1687 (1995); 
K.B. Davis {\it et al}., Phys. Rev. Lett. {\bf 75}, 3969 (1995).

\bibitem{molBEC}
M. Greiner, C.A. Regal, D.S. Jin, Nature {\bf 426}, 537 (2003);
S. Jochim {\it et al.}, Science {\bf 302}, 2101 (2003); 
M.W. Zwierlen {\it et al.}, Phys. Rev. Lett. {\bf 91}, 250401 (2003).

\bibitem{regal}

See, e.g., C.A. Regal, C. Ticknor, J.L. Bohn, and D.S. Jin, Nature {\bf 424}, 47 (2003).

\bibitem{durr}

See, e.g., S. D\"urr, T. Volz, A. Marte, and G. Rempe, Phys. Rev. Lett. {\bf 92}, 020406 (2004).

\bibitem{tiesinga}

E. Tiesinga, B.J. Verhaar, and H.T.C. Stoof, Phys. Rev. A {\bf 47}, 4114 (1993). 

\bibitem{inouye}

S. Inouye {\it et al.}, Nature {\bf 392}, 151 (1998).

\bibitem{kraemer}

T. Kraemer {\it et al.}, cond-mat/0512394.

\bibitem{FG}

V.V. Flambaum and J.S.M. Ginges, cond-mat/0207627.

\bibitem{stwalley}

W.C. Stwalley, Phys. Rev. Lett. {\bf 37}, 1628 (1976); 
Y.H. Uang, R.F. Ferrante, and W.C. Stwalley, J. Chem. Phys. 
{\bf 74}, 6267 (1981).

\bibitem{bosenova}

S.L. Cornish {\it et al}., Phys. Rev. Lett. {\bf 85}, 1795 (2000). 

\bibitem{PbO}
J.L. Bohn, A.V. Avdeenkov, and M.P. Deskevich, Phys. Rev. Lett. {\bf 89}, 203202 (2002).

\bibitem{Lane} 

A.M. Lane, R.G. Thomas, Rev. Mod.Phys. {\bf 30}, 257 (1958).

\bibitem{Gribakin}  V.V. Flambaum, A.A. Gribakina, G.F. Gribakin, and M.G. Kozlov, 
Phys. Rev. A {\bf 50}, 267 (1994); 
A.A. Gribakina, V.V. Flambaum, and G.F. Gribakin, Phys. Rev E {\bf 52}, 5667 (1995);
V.V. Flambaum, A.A. Gribakina, and G.F. Gribakin, Phys. Rev. A {\bf 54}, 2066 (1996); 
V.V. Flambaum, A.A. Gribakina, G.F. Gribakin, Phys. Rev. A {\bf 58}, 230 (1998);
G.F. Gribakin, A.A. Gribakina, and V.V. Flambaum, Aust. J. Phys. {\bf 52}, 443 (1999).

\bibitem{sushkov} 

O.P. Sushkov and V.V. Flambaum, Usp. Fiz. Nauk. {\bf 136}, 3 (1982) 
[Sov. Phys. Usp. {\bf 25}, 1 (1982)];
V.V. Flambaum and G.F. Gribakin, Prog. Part. Nucl. Phys. {\bf 35}, 423 (1995).  

\bibitem{pvreviews}

For general reviews on the possible origins of homochirality, 
see, e.g., 
M. Avalos {\it et al}., Tetrahedron: Asymmetry {\bf 11}, 2845 (2000); 
H. Buschmann, R. Thede, and D. Heller, Angew. Chem. Int. Ed. {\bf 39}, 4033 (2000). 

\bibitem{rein} 

D.W. Rein, J. Mol. Evol. {\bf 4}, 15 (1974).

\bibitem{letokhov}

V.S. Letokhov, Phys. Lett. {\bf 53A}, 275 (1975).

\bibitem{khriplovich}

I.B. Khriplovich, {\it Parity nonconservation in atomic phenomena} 
(Gordon and Breach, Philadelphia, 1991).

\bibitem{bouchiats}

M.-A. Bouchiat and C. Bouchiat, Rep. Prog. Phys. {\bf 60}, 1351 (1997).

\bibitem{RHS}

D.W. Rein, R.A. Hegstrom, and P.G.H. Sandars, Phys. Lett. {\bf 71A}, 499 (1979); 
R.A. Hegstrom, D.W. Rein, and P.G.H. Sandars, J. Chem. Phys. {\bf 73}, 2329 (1980). 

\bibitem{molcalcs}

J.K. Laerdahl and P. Schwerdtfeger, Phys. Rev. A {\bf 60}, 4439 (1999).

\bibitem{det} 

Ch. Daussy {\it et al.}, Phys. Rev. Lett. {\bf 83}, 1554 (1999).

\bibitem{four}

It is not necessary to have all four atoms different. 
Even such simple molecules as H$_2$O$_2$ have a chiral structure.

\bibitem{herbig}

J. Herbig {\it et al.}, Science {\bf 301}, 1510 (2003).

\bibitem{condon}

E.U. Condon, Rev. Mod. Phys. {\bf 9}, 432 (1937).

\bibitem{pola}

P.L. Polavarapu, Chirality {\bf 14}, 768 (2002).

\bibitem{moore}

M.G. Moore and A. Vardi, Phys. Rev. Lett. {\bf 88}, 160402 (2002).

\bibitem{Azimov} 

Ya.I. Azimov, A.A. Anselm, A.N. Moskalev, and R.M. Ryndin, Sov. Phys. JETP {\bf 40}, 8 (1975).

\bibitem{sf}

O.P. Sushkov and V.V. Flambaum, Pis'ma Zh. Eksp. Teor. Fiz. {\bf 32}, 377 (1980) 
[JETP Lett. {\bf 32}, 352 (1980)].

\bibitem{nNexp}

G.E. Mitchell, J.D. Bowman, and H.A. Widenmuller, Rev. Mod. Phys. {\bf 71}, 445 (1999).

\bibitem{ngamma} V.V. Flambaum and O.P. Sushkov, Nucl. Phys. A {\bf 435}, 352 (1985); 
Nucl. Phys. A  {\bf 412}, 13 (1984).

\end{thebibliography}
\end{document}